\documentclass[aip, cha,author-year, superscriptaddress,showpacs,floatfix]{revtex4-1}

\usepackage{epsfig}

\usepackage{textcomp}

\usepackage{array}
\usepackage{amssymb}
\usepackage{fancyhdr}

\newcommand{\be}{\begin{equation}}
\newcommand{\ee}{\end{equation}}
\newcommand{\bc}{\begin{center}}
\newcommand{\ec}{\end{center}}
\newcommand{\bi}{\begin{itemize}}
\newcommand{\ei}{\end{itemize}}
\newcommand{\ba}{\begin{eqnarray}}
\newcommand{\ea}{\end{eqnarray}}

\newcommand{\ignore}[1]{}

\usepackage{multirow}
\usepackage{amsmath}
\usepackage{graphicx}
\usepackage{color}
\usepackage{multirow}
\definecolor{Dark}{gray}{.90}

\begin{document}
\draft
\title{Enhanced repertoire of brain dynamical states during the psychedelic experience}
\author{Enzo Tagliazucchi}
\affiliation{Neurology Department and Brain Imaging Center, Goethe University, Frankfurt am 
Main, Germany}
\author{Robin Carhart-Harris}
\affiliation{Imperial College London, Centre for Neuropsychopharmacology,  Division of Experimental Medicine, London, UK}
\author{Robert Leech}
\affiliation{Computational, Cognitive and  Clinical Neuroimaging Laboratory (C3NL), Division of Brain Sciences, Imperial College 
London, UK}
\author{David Nutt}
\affiliation{Imperial College London, Centre for Neuropsychopharmacology,  Division of Experimental Medicine, London, UK}
\author{Dante R. Chialvo}
 \affiliation{Consejo Nacional de Investigaciones Cientificas y Tecnologicas (CONICET), 
Buenos Aires, Argentina.}
\date{\today}

\begin{abstract}
The study of rapid changes in brain dynamics and functional connectivity (FC) is 
of increasing interest in neuroimaging. Brain states departing from normal waking 
consciousness are expected to be accompanied by alterations in the aforementioned 
dynamics. In particular, the psychedelic experience produced by psilocybin (a substance 
found in ``magic mushrooms'') is characterized by unconstrained cognition and profound 
alterations in the perception of time, space and selfhood. Considering the spontaneous 
and subjective manifestation of these effects, we hypothesize that neural correlates 
of the psychedelic experience can be found in the dynamics and variability of spontaneous 
brain activity fluctuations and connectivity, measurable with functional Magnetic 
Resonance Imaging (fMRI). Fifteen healthy subjects were scanned before, during 
and after intravenous infusion of psilocybin and an inert placebo. Blood-Oxygen 
Level Dependent (BOLD) temporal variability was assessed computing the variance 
and total spectral power, resulting in increased signal variability bilaterally 
in the hippocampi and anterior cingulate cortex. Changes in BOLD signal spectral 
behavior (including spectral scaling exponents) affected exclusively higher brain 
systems such as the default mode, executive control and dorsal attention networks. 
A novel framework enabled us to track different connectivity states explored by 
the brain during rest. This approach revealed a wider repertoire of connectivity 
states post-psilocybin than during control conditions. Together, the present results 
provide a comprehensive account of the effects of psilocybin on dynamical behaviour 
in the human brain at a macroscopic level and may have implications for our understanding 
of the unconstrained, hyper-associative quality of consciousness in the psychedelic 
state.

\end{abstract}
\maketitle
\vspace{18pt}
\section{Introduction}

Psilocybin (phophoryl-4-hydroxy-dimethyltryptamine) is the phosphorylated ester 
of the main psychoactive compound found in magic mushrooms. Pharmacologically related 
to the prototypical psychedelic LSD, psilocybin has a long history of ceremonial 
use via mushroom ingestion and, in modern times, psychedelics have been assessed 
as tools to enhance the psychotherapeutic process (Grob et al., 2011; Krebs et 
al., 2012; Moreno et al., 2006). The subjective effects of psychedelics include 
(but are not limited to) unconstrained, hyper-associative cognition, distorted 
sensory perception (including synesthesia and visions of dynamic geometric patterns) 
and alterations in one's sense of self, time and space. There is recent preliminary 
evidence that psychedelics may be effective in the treatment of anxiety related 
to dying (Grob et al., 2011) and obsessive compulsive disorder (Moreno et al., 
2006) and there are neurobiological reasons to consider their potential as antidepressants 
(Carhart-Harris et al., 2012a,b). Similar to ketamine (another novel candidate 
antidepressant) psychedelics may also mimic certain psychotic states such as the 
altered quality of consciousness that is sometimes seen in the onset-phase of a 
first psychotic episode (Carhart-Harris et al., 2014). There is also evidence to 
consider similarities between the psychology and neurobiology of the psychedelic 
state and Rapid Eye Movement (REM) sleep (Carhart-Harris, 2007; Carhart-Harris 
\& Nutt, 2014), the sleep stage associated with vivid dreaming (Aserinsky and Kleitman, 
1953)

The potential therapeutic use of psychedelics, as well as their capacity to modulate 
the quality of conscious experience in a relatively unique and profound manner, 
emphasizes the importance of studying these drugs and how they act on the brain 
to produce their novel effects. One potentially powerful way to approach this problem 
is to exploit human neuroimaging to measure changes in brain activity during the 
induction of the psychedelic state. The neural correlates of the psychedelic experience 
induced by psilocybin have been recently assessed using Arterial Spin Labeling 
(ASL) and BOLD fMRI (Carhart-Harris et al., 2012a). This work found that psilocybin 
results in a reduction of both CBF and BOLD signal in major subcortical and cortical 
hub structures such as the thalamus, posterior cingulate (PCC) and medial prefrontal 
cortex (mPFC) and in decreased resting state functional connectivity (RSFC) between 
the normally highly coupled mPFC and PCC. Furthermore, our most recent study used 
magnetoencephalography (MEG) to more directly measure altered neural activity post-psilocybin 
and here we found decreased oscillatory power in the same cortical hub structures 
(Muthukumaraswamy et al., 2013, see also Carhart-Harris et al. 2014 for a review 
on this work). 

These results establish that psilocybin markedly affects BOLD, CBF, RSFC and oscillatory 
electrophysiological measures in strategically important brain structures, presumably 
involved in information integration and routing (Hagmann et al., 2008; Leech et 
al., 2012; de Pasquale et al., 2012; Carhart-Harris et al. 2014). However, the 
effects of psilocybin on the variance of brain activity parameters across time 
has been relatively understudied and this line of enquiry may be particularly informative 
in terms of shedding new light on the mechanisms by which psychedelics elicit their 
characteristic psychological effects. Thus, the main objective of this paper is 
to examine how psilocybin modulates the dynamics and temporal variability of resting 
state BOLD activity. 

Once regarded as physiological noise, a large body of research 
has now established that resting state fluctuations in brain activity have enormous 
neurophysiological and functional relevance (Fox and Raichle, 2007). Spontaneous 
fluctuations self-organize into internally coherent spatio-temporal patterns of 
activity that reflect neural systems engaged during distinct cognitive states (termed 
``intrinsic'' or ``resting state networks'' - RSNs) (Fox and Raichle, 2005; Raichle, 
2011; Smith et al., 2009). It has been suggested that the variety of spontaneous 
activity patterns that the brain enters during task-free conditions reflects the 
naturally itinerant and variegated quality of normal consciousness (Raichle, 2011). 
However, spatio-temporal patterns of resting state activity are globally well preserved 
in states such as sleep (Larson-Prior et al., 2009; Boly et al., 2009, 2012; Brodbeck 
et al., 2012; Tagliazucchi et al., 2013a,b,c) in which there is a reduced level 
of awareness  -- although very specific changes in connectivity occur across NREM 
sleep, allowing the decoding of the sleep stage from fMRI data (Tagliazucchi et 
al., 2012c; Tagliazucchi \& Laufs, 2014). Thus, if the subjective quality of consciousness 
is markedly different in deep sleep relative to the normal wakeful state (for example) 
yet FC measures remain largely preserved, this would suggest that these measures 
provide limited information about the biological mechanisms underlying different 
conscious states. Similarly, intra-RSN FC is decreased under psilocybin (Carhart-Harris 
et al., 2012b) yet subjective reports of unconstrained or even ``expanded'' consciousness 
are common among users (see Carhart-Harris et al. 2014 for a discussion). Thus, 
the present analyses are motivated by the view that more sensitive and specific 
indices are required to develop our understanding of the neurobiology of conscious 
states, and that measures which factor in variance over time may be particularly 
informative.

A key feature of spontaneous brain activity is its dynamical nature. In analogy 
to other self-organized systems in nature, the brain has been described as a system 
residing in (or at least near to) a critical point or transition zone between states 
of order and disorder (Chialvo, 2010; Tagliazucchi \& Chialvo, 2011; Tagliazucchi 
et al., 2012a; Haimovici et al., 2013). In this critical zone, it is hypothesized 
that the brain can explore a maximal repertoire of its possible dynamical states, 
a feature which could confer obvious evolutionary advantages in terms of cognitive 
and behavioral flexibility. It has even been proposed that this cognitive flexibility 
and range may be a key property of adult human consciousness itself (Tononi, 2012). 
An interesting research question therefore is whether changes in spontaneous brain 
activity produced by psilocybin are consistent with a displacement from this critical 
point - perhaps towards a more entropic or super-critical state (i.e. one closer 
to the extreme of disorder than normal waking consciousness) (Carhart-Harris et 
al. 2014). Further motivating this hypothesis are subjective reports of hyper-associative 
cognition under psychedelics, indicative of unconstrained brain dynamics. Thus, 
in order to test this hypothesis, it makes conceptual sense to focus on variability 
in activity and FC parameters over time, instead of the default procedure of averaging 
these over a prolonged period. 

In what follows, we present empirical data that tests the hypothesis that brain activity becomes less ordered in the psychedelic 
state and that the repertoire of possible states is enhanced. After the relevant 
findings have been presented, we engage in a discussion to suggest possible strategies 
that may further characterize quantitatively where the ``psychedelic brain'' resides 
in state space relative to the dynamical position occupied by normal waking consciousness.

\section{MATERIALS AND METHODS}

\textbf{Study overview and design}
\vspace{10pt}

This was a within-subjects placebo-controlled study. The study was approved by 
a local NHS Research Ethics Committee and University of Bristol Research and Development 
department, and conducted in accordance with Good Clinical Practice guidelines. 
A Home Office License was obtained for storage and handling of a Schedule 1 drug 
and the University of Bristol sponsored the research.

\vspace{10pt}
\textbf{Participants}
\vspace{10pt}

This is a new analysis on previously published data (Carhart-Harris et al., 2012a,b). 
Fifteen healthy subjects took part: 13 males and 2 females (mean age = 32, SD = 
8.9). Recruitment was via word of mouth. All subjects were required to give informed 
consent and undergo health screens prior to enrollment. Study inclusion criteria 
were: at least 21 years of age, no personal or immediate family history of a major 
psychiatric disorder, substance dependence, cardiovascular disease, and no history 
of a significant adverse response to a hallucinogenic drug. All of the subjects 
had used psilocybin at least once before (mean number of uses per subject = 16.4, 
SD = 27.2) but not within 6 weeks of the study.

\vspace{10pt}
\textbf{Anatomical Scans}
\vspace{10pt}

Image acquisition was performed on a 3T GE HDx system. Anatomical scans were performed 
before each functional scan and thus during sobriety. These were 3D fast spoiled 
gradient echo scans in an axial orientation, with field of view = 256\ensuremath{\times}256\ensuremath{\times}192 
and matrix = 256\ensuremath{\times}256\ensuremath{\times}192 to yield 1 mm isotropic 
voxel resolution (repetition time/echo time [TR/TE] = 7.9/3.0 ms; inversion time 
= 450 ms; flip angle = 20°).

\vspace{10pt}
\textbf{Drug infusion}
\vspace{10pt}

All subjects underwent two 12 min eyes closed resting state BOLD fMRI scans on 
2 separate occasions at least 7 days apart: placebo (10 ml saline, 60 s intravenous 
injection) was given on one occasion and psilocybin (2 mg dissolved in 10 ml saline, 
60 s infusion) on the other. Seven of the subjects received psilocybin in scan 
1, and 8 received it in scan 2. Injections were given manually by a doctor within 
the scanning suite. The 60 s infusions began exactly 6 min after the start of the 
12 min scans. The subjective effects of psilocybin were felt almost immediately 
after injection and were sustained for the remainder of the RS scan. For more details 
on the subjective effects of intravenous psilocybin see (Carhart-Harris et al., 
2012a,b)

\vspace{10pt}
\textbf{fMRI Data acquisition and pre-processing}
\vspace{10pt}

BOLD-weighted fMRI data were acquired using a gradient echo planar imaging sequence, 
TR/TE 3000/35 ms, field-of-view = 192 mm, 64 \ensuremath{\times} 64 acquisition 
matrix, parallel acceleration factor = 2, 90° flip angle. Fifty-three oblique 
axial slices were acquired in an interleaved fashion, each 3 mm thick with zero 
slice gap (3 \ensuremath{\times} 3 \ensuremath{\times} 3 mm voxels). A total of 
240 volumes were acquired, with infusion taking place in the middle of the session. 
Data were motion corrected using FSL MCFLIRT and a high-pass filter of 100 s was 
applied. Data were smoothed using a Gaussian kernel of 5 mm FWHM. Motion time courses 
were regressed from the data during the pre-processing step, together with average 
CSF and white matter time courses. As an additional control to exclude epochs of 
large movement amplitude, all volumes associated with a mean head displacement 
larger than 1 s.d. were erased from the analysis. 

\vspace{10pt}
\textbf{Analysis of spatio-temporal dynamics}
\vspace{10pt}

The methods here employed are based on a statistical physics framework useful for 
characterizing fluctuations in systems composed of large number of coupled degrees 
of freedom. From this perspective, fluctuating activity in a particular region 
and correlations between different regions are interdependent or related (Ross, 
1966). Therefore, any change in the cortical dynamics due to a given intervention 
is expected to be reflected both at the level of the variance of the activity at 
one region and in the strength of the interactions between different regions. In 
what follows, the methodology accounting for these two aspects (i.e. the temporal 
variability and the dynamical changes in correlations) are explained in detail. 
All the numerical analyses for these calculations were performed using in-house 
MATLAB scripts.

\vspace{10pt}
\textbf{Analysis of temporal variability}
\vspace{10pt}

The analysis referred to here as ``temporal variability'' is concerned with the 
variance in the amplitude of the BOLD signal and it can be expressed both in a 
temporal (i.e., the standard variance) and frequency domain, as will be explained 
below. As a time domain measure of variability in the amplitude of the BOLD signal, 
straightforward voxel-wise computation of BOLD variance was performed, resulting 
in whole-brain BOLD variance maps for each condition (psilocybin/placebo before 
and after infusion) and participant. An additional evaluation of variance can be 
performed in the frequency domain, noting that by Parseval's theorem:

\begin{equation}
\int x^2(t)dt = \int |A(f)|^2df
\end{equation}

Then, the following series of equalities holds (for a signal with zero mean):

\begin{equation}
\sigma^2 = \lim_{T \to \infty} \frac{1}{T} \int_{-T/2}^{T/2} x^2(t)dt =  \lim_{T \to \infty} \frac{1}{T} \int  |A(f)|^2df = \int \Phi(f)df
\end{equation}

In Eqs. 1 and 2, A(f) represents the Fourier transform of x(t) and $\Phi$(f) is 
the power spectral density. Therefore, the variance can also be obtained by integrating 
$\Phi$(f) across the whole range of frequencies.

\vspace{10pt}
\textbf{Further spectral analysis of BOLD fluctuations}
\vspace{10pt}

Eqs. 1 and 2 show that the integral of the power spectral density equals the variance 
of the signal. If this integral is not performed across all frequencies but in 
a certain range, the result will correspond to the contribution of the frequencies 
to the variance in the specified ranges. It is generally considered that slow (0.01 
- 0.1 Hz) BOLD frequencies carry neural significance in the resting state fMRI 
signal (Cordes et al., 2001), therefore, we evaluated separately the low frequency 
power (LFP) in this range and compared it before and after the psilocybin infusion.

Given the fact that the spectral content of spontaneous BOLD fluctuations, like 
many other complex systems in nature, follows a power law of the form $\frac{1}{f^{\alpha}}$ (Expert et al., 2011), its power spectrum density can also be characterized by 
a single parameter ($\alpha$). This parameter condenses the scaling behavior and 
is demonstrative of the long-range temporal correlations (or memory) of any given 
signal. Thus, for uncorrelated noise, $\alpha$ = 0 (i.e. a flat spectrum, or so 
called ``white noise'') is obtained, whereas for signals presenting some degree 
of long-term correlations, values of $\alpha$ $>$ 0 (i.e., the so called 
``colored noise'') are observed. To obtain the scaling exponent $\alpha$, the first 
derivative of BOLD time series was first computed (in order to minimize the influence 
of non-stationarities) and the fit was performed on the spectrum of the derivative. 
Note that for a power spectrum of the form $\frac{1}{f^{\alpha}}$, differentiation decreases the value 
of  by 2, so the exponent of the original time series can be recovered from that 
of the derivative.

\vspace{10pt}
\textbf{Point-process analysis}
\vspace{10pt}

A recent series of studies demonstrated that the continuous BOLD signal can be 
transformed into a discrete point-process encoding the timings of the most functionally 
relevant events (Tagliazucchi et al., 2010a,b, 2012a; Petridou et al., 2012, Davis 
et al., 2013). In this approach, relevant events are defined by a threshold crossing 
(e.g. whenever the signal departs +1 s.d. from its mean value). Besides allowing 
a dramatic compression of the information present in a BOLD timeseries, this approach 
has the particular benefit of allowing a simple criterion to eliminate motion artifacts: 
i.e. ``points'' (threshold crossings) occurring during high movement epochs (i.e. 
larger than a certain value) are simply ignored in the analysis.

Two interdependent observables are defined once the point-process is obtained: 
the rate (number of crossings divided by the series length) and inter-event intervals 
(average temporal separation between two consecutive points). On average, there 
is a clear inverse relationship between these two variables. Furthermore, the rate 
is expected to increase (or the interval to decrease) for a signal with a high 
contribution of fast frequencies.

As an efficient alternative to the spectral analysis described in the previous 
section, we computed the voxel-wise distribution of rates and intervals and compared 
it between the different conditions.

\vspace{10pt}
\textbf{Computation of dynamical functional connectivity states and its associated 
entropy}
\vspace{10pt}

As already discussed above, for spatio-temporal fluctuations arising from the dynamics 
of large scale systems, the relation between the temporal fluctuations of the average 
signal and its spatial correlation function is well defined (Ross, 1966). It is 
known that under general assumptions, the mean field peak-to-peak amplitude of 
a signal (or its variance) is directly proportional to its mean correlation value. 
Intuitively, this is analogous to the principle that synchronized/desynchronized 
clapping produces stronger/weaker collective effects. Therefore, to complement 
the investigation of regional changes in the variance of the BOLD signal in various 
regions of the cortex, it is logical to also look at measures of the mean functional 
correlations between participating regions. Given the highly transient nature of 
these correlations, we term these indices ``dynamical functional connectivity states''. 
They are computed as follows (see Fig. 4 for an illustration of the general procedure).

The BOLD time series is first divided into M non-overlapping windows of length 
L. Then, the partial correlation values between a set of N brain regions of interest 
is computed inside each temporal window as follows,

\begin{equation}
R_C(X,Y) = \min R(X,Y | Z)
\end{equation}
where $R(X,Y|Z)$ is defined as follows,
\begin{equation}
R(X,Y|Z)  =  \frac{R(X,Y)-R(Y,Z)}{ \sqrt{1-R(X,Z)^2}}
\end{equation}

In Eq. 3, the minimum is taken across all signals Z different from X,Y. In Eq. 
4, R(X,Y) represents the linear correlation between variables X and Y. Thus, the 
partial correlation $R_{C}(X,Y)$ measures the correlation of both variables 
removing the effect of a set of controlling variables. In what follows, the set 
of variables will include all BOLD signals from the N brain regions under study 
plus the time series of absolute head displacement (to remove spurious correlations), 
as estimated during the motion correction step. Finally, for each pair of regions 
a link is established between them if the correlation p-value is significant at 
the level of p $<$ 0.05, Bonferroni corrected $(N (N - 1)/2)$. Performing 
this computation for all temporal windows gives the temporal evolution of the connectivity 
graph.

For N brain regions, the total number of dynamical functional connectivity graphs 
(i.e. all possible connectivity motifs) equals $2^{(N(N-1)/2)}$. 
For example, for N = 3, 4, 5, the number of states is 8, 64, 1024. Once the dynamical 
functional connectivity states are obtained by the procedure above described, a 
symbolization procedure can be used to map the limited amount of states into discrete 
symbols (a bijection or one-to-one mapping of each temporal sequence of states 
into ``words'' - whose letters represent the different connectivity graphs). Then, 
the entropy of this sequence can be computed using the straightforward definition 
of Shannon's entropy: 

\begin{equation}
H = \sum_{i=1}^{N(N-1)/2} p_i \log(1/p_i)
\end{equation}

where $p_{i }$ is the probability of the i-th state occurring in 
the sequence.

We evaluate the entropy of dynamical functional connectivity states 
in the network of regions of peak statistical significance presented in Table 1 
and Fig. 1 (i.e., the left and right hippocampus and ACC). This choice is motivated 
by the well-known observation that, for a system of coupled elements, the temporal 
variability of the activity time series is driven by variability in the collective 
interactions (Ross, 1966). This process resulted in 4 regions and a total of 64 
connectivity states (i.e. the number of possible dynamical interaction motifs between 
the 4 nodes).

\vspace{10pt}
\textbf{Statistical testing}
\vspace{10pt}

To perform statistical significance testing, a paired t-test was used, as implemented 
in SPM8. Only clusters passing a threshold of p $<$ 0.05 Family Wise Error 
(FWE) corrected for multiple comparisons were considered significant. For display 
purposes, significance maps were thresholded at p $<$ 0.005, showing only 
clusters passing the above mentioned criterion. Significance testing for the entropy 
differences reported in Fig. 5 was also evaluated using a paired t-test.

\section{RESULTS}

The statistical significance maps for the BOLD signal variance and total spectral 
power can be found in Fig. 1A and the significance peaks are summarized in Table 
1 (in this and the following sections, unless explicitly stated no differences 
were found by examining the opposite contrasts). Both measures show increased variability 
following psilocybin administration - both in the temporal and spectral domain 
- with peaks in the anterior cingulate cortex and bilateral hippocampus (total 
spectral power is also increased in the bilateral parahippocampal gyri). As shown 
in Fig. 1 there is a large overlap in the regions affected by both measures, as 
can be expected from Eq. 2. The differences ~found ~for the power spectral estimation 
in the parahippocampal gyrus are not apparent for the variance. Most likely this 
discrepancy is due to the numerical evaluation of these two equivalent quantities.

To find the temporal evolution of the variance increases, we selected the four 
significance peaks for BOLD variance as ROIs (Table 1) and used a sliding window 
analysis (1 min. window length) to compute the evolution of the variance. Since 
these ROI are selected as regions of high variance, it is already known that fluctuations 
in their activity will have an increased variance relative to baseline. However, 
this analysis provides additional information about when those increases occurred 
(e.g. just after the psilocybin infusion or later). In Fig. 1B the results of this 
analysis are presented. A homogeneous increase in signal variance is observed during 
the first 3 minutes post-psilocybin infusion, with a tendency to plateau or slightly 
decrease afterwards. A longer experiment would be needed to confirm the persistence 
of this effect, but it does seem to correspond well with the known pharmaco-dynamics 
of intravenous psilocybin, i.e. subjects reported the most intense subjective effects 
within the first 3 minutes of the infusion and the effects were persistent for 
the duration of the scan (Carhart-Harris et al., 2012a,b).

We also studied the variance of intra-hippocampal connectivity time courses, computed 
as the linear correlation of all hippocampal voxels over non-overlapping windows 
of different lengths. Results are shown in Fig. 1C. It can be seen that the variance 
is higher in both hippocampi after the psilocybin infusion.

Next, we studied the spectral content of spontaneous BOLD fluctuations. This was 
done by computing the low frequency (0.01-0.1 Hz) power (LFP) and the power spectrum 
scaling exponent ($\alpha$). Statistical significance maps are presented in Fig. 
2A and statistical significance peaks in Table 2.

After psilocybin infusion, diffuse widespread decreases in LFP and the scaling 
exponent $\alpha$ were observed in frontal and parietal regions. Changes in LFP 
and the scaling exponent ($\alpha$) were found consistently in the same spatial 
locations, which is expected since a scaling closer to that of uncorrelated noise 
(i.e. an $\alpha$ value closer to 0) will result in weaker low frequency spectral 
power. These effects were confined to the post-psilocybin period. No differences 
were found when performing the same comparison in the placebo condition.

To further characterize BOLD fluctuations, we transformed whole brain signals into 
a spatio-temporal point-process and extracted two statistics: the rate (average 
number of events) and the interval (average separation between two events). By 
the very definition of the point-process, a signal with high power in fast frequencies 
will give a high rate and therefore a low separation between events i.e., the average 
interval will be small. Results shown in Fig. 2B confirm this observation. Statistical 
significance peaks are presented in Table 3. A rate increase and interval decrease 
after psilocybin was observed in parietal and frontal regions largely overlapping 
with those of Fig. 2A. No differences were found when performing the same comparison 
in the placebo condition.

To identify the brain networks associated with the changes shown in Figs. 2A and 
2B, we computed the overlap of the statistical significance maps with a set of 
well-established Resting State Networks (RSNs) (Beckmann and Smith, 2004; Beckmann 
et al., 2005). These included: two visual (medial and lateral), an auditory, sensori-motor, 
default mode, executive control and two lateralized dorsal attention networks.

To compute the overlap for a given map, we counted the number of voxels included 
in each RSN and normalized by the total number of voxels in the RSN mask. Using 
this approach, larger maps present a higher chance of having large overlaps, therefore, 
we constructed a null-hypothesis by randomizing the phases of the maps (after transforming 
to Fourier space), resulting in images with the same second order statistics (see 
also Tagliazucchi et al., 2013c). A total of 100 randomizations were performed 
for each comparison and an empirical p-value was constructed, counting the ratio 
of instances in which the real overlap exceeded the overlap computed with the randomized 
map. The results of these analyses are presented in Fig. 3. For LFP and $\alpha$, 
significant overlaps were detected in the default mode, control and attention networks, 
whereas all sensory networks remained unaffected. For the point-process rate (PPR) 
and point-process interval (PPI) the changes were locally confined to the default 
mode network.

Next, the dynamic functional connectivity states were obtained from the set of 
regions showing increased temporal variability (Fig. 1 and Table 1) by applying 
the analysis outlined in Fig. 4. Using different window lengths (ranging from 15 
to 150 s.) we evaluated the entropy of the distribution of connectivity states 
in the network comprised by two ACC ROIs and the bilateral hippocampi. An entropy 
increase was found when comparing the results between the periods before and after 
the psilocybin infusion and after the psilocybin infusion vs. after the placebo 
infusion, but no changes were observed when comparing before and after placebo 
(see Fig. 5A). The entropy increase was not seen for very short window lengths 
but was manifest for all lengths larger than approximately 1 min.

The most frequent states in each condition are shown in Fig. 5B. These have sparse 
connectivity and represent functional connections between homologous ROIs and between 
ROIs in the same hemisphere, but cross-hemisphere connections between hippocampal 
and ACC ROIs also appear in the psilocybin condition. The psilocybin state is also 
characterized by a larger repertoire of states: i.e. novel motifs that are exclusive 
to the psychedelic state and which are shown in the last row of Fig. 5B. These 
motifs are among the most interconnected states possible.

\section{DISCUSSION}

The novel analyses described in this paper provide a considerable amount of new 
information about how psilocybin affects brain activity. In summary, increased 
variance in the BOLD signal was observed in the bilateral hippocampi and ACC and 
the temporal dynamics of these increases corresponded well with the rapid pharmacodynamics 
of intravenous psilocybin (Carhart-Harris et al., 2012a,b). Decreased low frequency 
power and frequency scaling exponent (indicative of a less correlated signal) were 
observed in higher-level association regions accompanied by an increase in the 
point process rate at the default mode network. Perhaps the most novel and intriguing 
aspect of the present analyses was our assessment of dynamical functional connectivity 
within a simple hippocampal/ACC network. Specifically, a greater diversity of connectivity 
motifs was observed after psilocybin, reflecting increased entropy in this system's 
dynamical behavior. Overall these quantities demonstrate an increase in the dynamical 
repertoire (i.e., new states) in the brain under psilocybin as well as an increase 
in the rate at which the repertoire is examined.

We first interpret the changes described in the results section from a purely dynamical 
point of view. The results presented in Fig. 1 represent a larger variance in the 
``mean field activity'' of the hippocampus and ACC. This change in variance is 
the expression of an increased amplitude of the BOLD signal fluctuations in these 
regions. From a statistical physics point of view, increased signal amplitude implies 
increased synchronization and so increased correlated activity in the source structures. 
This is also reflected in the analysis of the repertoire of connectivity states 
in a hippocampal-ACC network (discussed in Figs. 4 and 5). These two results together 
reveal an increased variability in the collective repertoire (i.e. a larger number 
of motifs) of metastable states (Tognoli \& Kelso, 2014) within the psychedelic 
state.

The increased amplitude fluctuations in the hippocampus are particularly intriguing 
given early depth electroencephalography (EEG) work that recorded similar abnormalities 
in hippocampal activity after LSD and mescaline (Schwarz et al., 1956; Monroe and 
Heath, 1961). Similar increases in oscillation amplitude have also been observed 
in the hippocampus using depth EEG recordings in patients showing nondrug-induced 
psychotic symptoms (Sem-Jacobsen et al., 1956; Heath, 1954) and bursts of high 
amplitude activity have been seen in human rapid-eye movement (REM) sleep (Cantero 
et al., 2003). Moreover, other imaging modalities have implicated increased hippocampal 
activity in psychosis (e.g., Friston et al., 1992) and REM sleep (Braun et al., 
1997; Miyauchi et al., 2009). Given that phenomenological similarities have previously 
been noted between the psychedelic, psychotic and dream states (Carhart-Harris, 
2007; Carhart-Harris \& Nutt, 2014), it is intriguing to consider whether altered 
hippocampal activity may be an important common property of these states. 

It has long been claimed that the psychedelic (translated ``mind-revealing'' (Huxley 
et al., 1977)) state is an expanded state of consciousness in which latent psychological 
material can emerge into consciousness (Cohen, 1967) and novel associations can 
form. Indeed, this was the original rationale for the use of LSD in psychotherapy 
(Busch and Johnson, 1950). It has also been claimed that psychedelics may be able 
to assist the creative process, for example, by promoting divergent thinking and 
remote association (Fadiman, 2011). Thus, the increased repertoire of metastable 
states observed here with psilocybin may be a mechanism by which these phenomena 
occur (see also Carhart-Harris et al. 2014)

It was also interesting that under psilocybin, more interhemispheric dynamical 
correlations were detected in the hippocampal/ACC network (Fig 5C). Recent electrophysiological 
work in mice has shown that layer 5 pyramidal neurons (the primary cellular units 
implicated in the action of psychedelic drugs (Muthukumaraswamy et al., 2013)) 
that are sensitive to serotonin 2A receptor mediated excitation (the primary pharmacological 
process implicated in the action of psychedelics (Muthukumaraswamy et al., 2013)) 
are disproportionately those pyramidal neurons that project interhemispherically 
(Avesar and Gulledge, 2012). Thus, altered interhemispheric communication may be 
an important component of the mechanism of action of psychedelics.

Concerning the low frequency fluctuations results, these were consistent with our 
earlier work with MEG in which we observed decreased oscillatory power in neural 
fields in high-level cortical regions in the 1-100 Hz frequency range (Muthukumaraswamy 
et al., 2013). In the present analysis, decreased low frequency power in the .01-.1Hz 
range was found and again, these effects were localized to consistent high-level 
cortical regions. Low frequency fluctuations in BOLD are known to correlate with 
neuronal parameters such as fluctuating gamma power and infraslow fluctuations 
in local field potentials (Pan et al., 2013). The slower beta band also shows positive 
correlations with fMRI fluctuations in key DMN regions (Laufs et al., 2003), whereas 
both alpha and beta apparently inhibit large-scale cortical BOLD coherence (Tagliazucchi 
et al., 2012b). Thus, it seems that a primary action of psilocybin, and likely 
other psychedelics (Riba et al., 2002), is to cause a generalized desynchrony and 
loss of oscillatory power in higher level cortical regions - likely via serotonin 
2A receptor mediated excitation of deep-layer pyramidal neurons in these regions 
(Muthukumaraswamy et al., 2013)). However, the high amplitude activity detected 
in the hippocampi and ACC (Fig. 1) suggests that this desynchronizing effect does 
not generalize to these deeper structures.

The frequency domain result was further examined by a separate analysis of the 
point-process rate and interval distributions (see Fig. 3). The RSNs which exhibited 
the most significant changes correspond to higher brain systems such as the DMN, 
executive control and attention networks and not primary sensory and motor networks. 
This outcome is consistent with the regional distribution of serotonin 2A receptors 
(Erritzoe et al., 2009), the receptors implicated in psilocybin's psychedelic action 
(Vollenweider et al., 1998). These receptors are concentrated in higher level cortical 
regions (e.g. the highest distribution in humans is in the PCC (Carhart-Harris 
et al., 2012b; Erritzoe et al., 2009) and are relatively less prevalent in the 
sensori-motor cortex. That the default mode network has consistently been implicated 
in the action of psilocybin is also intriguing given its association with self-reflection 
(Gusnard et al., 2001) and selfhood more generally (Carhart-Harris and Friston, 
2010). It is likely to be relevant therefore that one of the most commonly reported 
features of an intense psychedelic experience is a compromised sense of selfhood 
typically described as ``ego dissolution'' or ``ego disintegration'' (Klee, 1963; 
Savage, 1955; Carhart-Harris et al. 2014).

A potential limitation of our study arises from the possibility that psilocybin 
modifies the coupling between neuronal sources and hemodynamic activity as measured 
with fMRI (i.e. the hemodynamic response function). However, we note that there 
exists no evidence for this so far and, on the other hand, there exists ample direct 
electrophysiological evidence showing that psilocybin modifies brain activity in 
a way compatible with our findings (Muthukumaraswamy et al., 2013).

Another limitation stems from the limited number of regions included in the definition 
of dynamical states. This limited number is required to perform an exhaustive counting 
of all possible states, however, differences in the motifs between conditions (Fig. 
5) could arise by direct influence of regions outside the scope of the analysis, 
which are directly wired to a pair of nodes. Future work should address this limitation 
and attempt to track those regions to obtain a more complete picture of how dynamical 
states differ after and before psilocybin infusion.

Finally, it is important to offer some cautionary notes on the analyses documented 
in this paper. Mapping relatively transient dynamical motifs based on resting state 
fMRI data is a new and exploratory technique. Thus, further work is required to 
clarify its functional meaning. More specifically, given the Zipf-like behavior 
of the state distribution probability (Fig. 5B), states that appear rarely (i.e. 
at the tail of a power law distribution) cannot be trivially tested using the usual 
statistical significance, i.e. they do not fit easily into the standard frequentist 
framework of t-test statistics, in contrast to well established functional connectivity 
analyses that represent an average over prolonged periods of time. Thus, further 
mathematical testing and empirical work is required before we can begin to make 
confident predictions about the functional relevance of outcomes derived from their 
application.

\section{ACKNOWLEDGMENTS}

Work supported by CONICET (Argentina) and LOEWE Neuronale Koordination Forschungsschwerpunkt 
Frankfurt - NeFF (Germany). The original experiments were performed as part of 
a broader Beckley-Imperial psychedelic research programme. Authors thank Christian 
Beckmann for sharing the RSN masks reported in Beckmann et al. (2005). \pagebreak{}


\begin{tabular}
{|>{\raggedright}p{88pt}|>{\raggedright}p{104pt}|>{\raggedright}p{66pt}|>{\raggedright}p{43pt}|}
\hline
\textbf{MNI coordinates} & \textbf{AAL region} & \textbf{Hemisphere} & \textbf{t-value}\tabularnewline
\hline
\multicolumn{4}{|p{302pt}|}{BOLD variance (after psilocybin $>$before psilocybin)}\tabularnewline
\hline
(-34, -22, -16) & Hippocampus & Left & 4.51\tabularnewline
\hline
(26, -22, -16) & Hippocampus & Right & 3.54\tabularnewline
\hline
(-2, 22, 28) & Anterior cingulate & Left & 3.72\tabularnewline
\hline
(4, 34, 18) & Anterior cingulate & Right & 3.74\tabularnewline
\hline
\multicolumn{4}{|p{302pt}|}{BOLD total spectral power (after psilocybin \texttt{>} 
before psilocybin)}\tabularnewline
\hline
(32, -14, -13) & Hippocampus & Right & 4.03\tabularnewline
\hline
(-26, -18, -15) & Hippocampus & Left & 4.39\tabularnewline
\hline
(-26, 1, 38) & Parahippocampal Gyrus & Left & 4.77\tabularnewline
\hline
(37, -2, -38) & Parahippocampal gyrus & Right & 4.55\tabularnewline
\hline
(7, 27, 17) & Anterior cingulate & Right & 4.02\tabularnewline
\hline
(-7, 43, -2) & Anterior cingulate & Left & 4.12\tabularnewline
\hline
\end{tabular}

\textbf{Table 1:} Regions corresponding to local maxima of statistical significance (provided they are more than 8 mm apart) for increased cortical BOLD variance and total spectral power after psilocybin infusion (compared to before psilocybin infusion). Montreal Neurological Institute (MNI) coordinates and Automatic Anatomical Labeling (AAL) regions are provided.

\vspace{24pt}

\begin{tabular}{|>{\raggedright}p{88pt}|>{\raggedright}p{120pt}|>{\raggedright}p{54pt}|>{\raggedright}p{39pt}|}
\hline
\textbf{MNI coordinates} & \textbf{AAL region} & \textbf{Hemisphere} & \textbf{t-value}\tabularnewline
\hline
\multicolumn{4}{|p{302pt}|}{BOLD low frequency power (after psilocybin \texttt{<} 
before psilocybin)}\tabularnewline
\hline
(11, -60, 40) & Precuneus & Right & 4.92\tabularnewline
\hline
(38, -67, 44) & Angular gyrus & Right & 4.60\tabularnewline
\hline
(-34, -69, 44) & Angular gyrus & Left & 5.05\tabularnewline
\hline
(-39, -42, 42) & Inferior parietal cortex & Left & 4.66\tabularnewline
\hline
(-39, 45, -3) & Middle frontal gyrus & Left & 5.10\tabularnewline
\hline
(47, 44, -3) & Inferior frontal gyrus & Right & 5.08\tabularnewline
\hline
(5, 56, -9) & Middle frontal gyrus (orbital) & Right & 4.28\tabularnewline
\hline
\multicolumn{4}{|p{302pt}|}{BOLD power spectrum scaling exponent $\alpha$ (after 
psilocybin \texttt{<} before psilocybin)}\tabularnewline
\hline
(4,-59,33) & Precuneus & Right & 4.86\tabularnewline
\hline
(53,-61,33) & Angular gyrus & Right & 4.31\tabularnewline
\hline
(-48,-55,34) & Angular gyrus & Left & 4.20\tabularnewline
\hline
(-40,-34,60) & Postcentral gyrus & Left & 4.80\tabularnewline
\hline
(6,-22,60) & Paracentral lobule & Right & 4.79\tabularnewline
\hline
(50,-30,60) & Superior parietal cortex & Right & 5.73\tabularnewline
\hline
(-41,14,33) & Middle frontal gyrus & Left & 4.32\tabularnewline
\hline
(52,7,28) & Precentral gyrus & Right & 5.12\tabularnewline
\hline
(38,41,28) & Middle frontal gyrus & Right & 6.22\tabularnewline
\hline
(-27,43,28) & Middle frontal gyrus & Left & 5.10\tabularnewline
\hline
(10,55,-6) & Middle frontal gyrus (orbital) & Right & 4.48\tabularnewline
\hline
\end{tabular}

\vspace{24pt}
\textbf{Table 2}: Regions corresponding to local maxima of statistical significance 
(more than 8 mm apart) for decreased cortical BOLD low frequency 
power and power spectrum scaling exponent after psilocybin infusion (compared to 
before psilocybin infusion).

\pagebreak
 
 \begin{tabular}{|>{\raggedright}p{88pt}|>{\raggedright}p{104pt}|>{\raggedright}p{66pt}|>{\raggedright}p{43pt}|}
\hline
\textbf{MNI coordinates} & \textbf{AAL region} & \textbf{Hemisphere} & \textbf{t-value}\tabularnewline
\hline
\multicolumn{4}{|p{302pt}|}{Point-process rate (after psilocybin \texttt{>} before 
psilocybin)}\tabularnewline
\hline
(5, -52, 25) & Precuneus & Right & 4.47\tabularnewline
\hline
(47, -66, 29) & Angular gyrus & Right & 3.99\tabularnewline
\hline
(-41, -66, 24) & Angular gyrus & Left & 2.84\tabularnewline
\hline
(-41, 47, -5) & Middle frontal gyrus (orbital) & Left & 2.82\tabularnewline
\hline
(-41, 41, 24) & Middle frontal gyrus & Left & 2.97\tabularnewline
\hline
(35, 40, 24) & Middle frontal gyrus & Right & 3.14\tabularnewline
\hline
\multicolumn{4}{|p{302pt}|}{Point-process interval (after psilocybin \texttt{<} 
before psilocybin)}\tabularnewline
\hline
(5, -57, 25) & Precuneus & Right & 4.06\tabularnewline
\hline
(50, -58, 36) & Angular gyrus & Right & 3.98\tabularnewline
\hline
(-34, -66, 36) & Angular gyrus & Left & 3.00\tabularnewline
\hline
(-37, -32, 62) & Postcentral gyrus & Left & 3.35\tabularnewline
\hline
(41, -39, 62) & Paracentral lobule & Right & 3.18\tabularnewline
\hline
(-26, -9, 60) & Postcentral gyrus & Left & 3.59\tabularnewline
\hline
(28, 17, 54) & Middle frontal gyrus & Right & 3.72\tabularnewline
\hline
(-20, 15, 54) & Superior frontal gyrus & Left & 2.87\tabularnewline
\hline
(-22, 59, 12) & Superior frontal gyrus & Left & 3.54\tabularnewline
\hline
(4, 54, 6) & Anterior cingulate & Right & 3.29\tabularnewline
\hline
\end{tabular}

\vspace{12pt}
\textbf{Table 3:} Regions corresponding to local maxima of statistical significance 
(provided they are more than 8 mm apart) for increased cortical point-process rate 
(PPR) and decreased point-process interval (PPI) after psilocybin infusion (compared 
to before psilocybin infusion).

\pagebreak{}

\begin{figure}[ht]
\centering \psfig{figure=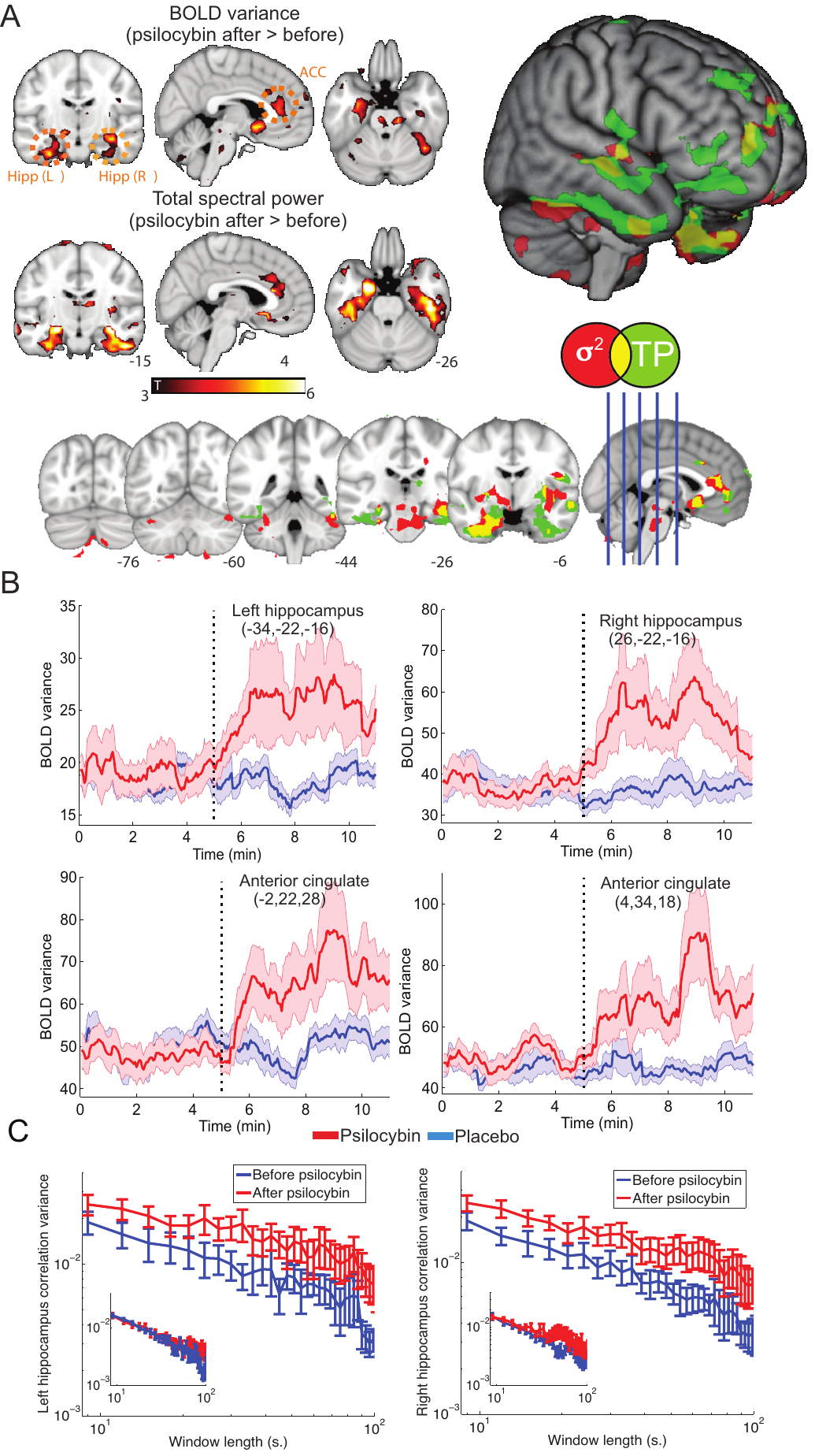,width= 4 truein,clip=true,angle=0} 
\caption{Psilocybin infusion modifies the temporal variability of BOLD  signal in a network comprising anterior cingulate cortex (ACC) and bilateral hippocampus. 
A) Maps of statistical significance for variance ($\sigma$\textsuperscript{2}) 
and total spectral power (TP) increases after psilocybin infusion. Results are 
shown overlaid separately into an anatomical MNI152 template, overlaid together 
and also rendered together into a 3D anatomical image. In all cases only clusters 
surviving a threshold of p $<$ 0.05, Family Wise Error (FWE) cluster corrected 
(after passing an uncorrected threshold of p $<$ 0.005) are shown. B) BOLD 
Variance time courses (obtained over a 1 min. sliding window) into the four regions 
of peak statistical significance defined in Table 1 for the psilocybin and the 
placebo infusion. C) Variance of the time course of intra-hippocampal correlations 
(computed over a range of non-overlapping window lengths) before and after psilocybin 
infusion. Results for placebo infusion are shown as insets (before and after in 
blue and red, respectively).}
\end{figure} 

\pagebreak{}

\begin{figure}[ht]
\centering \psfig{figure=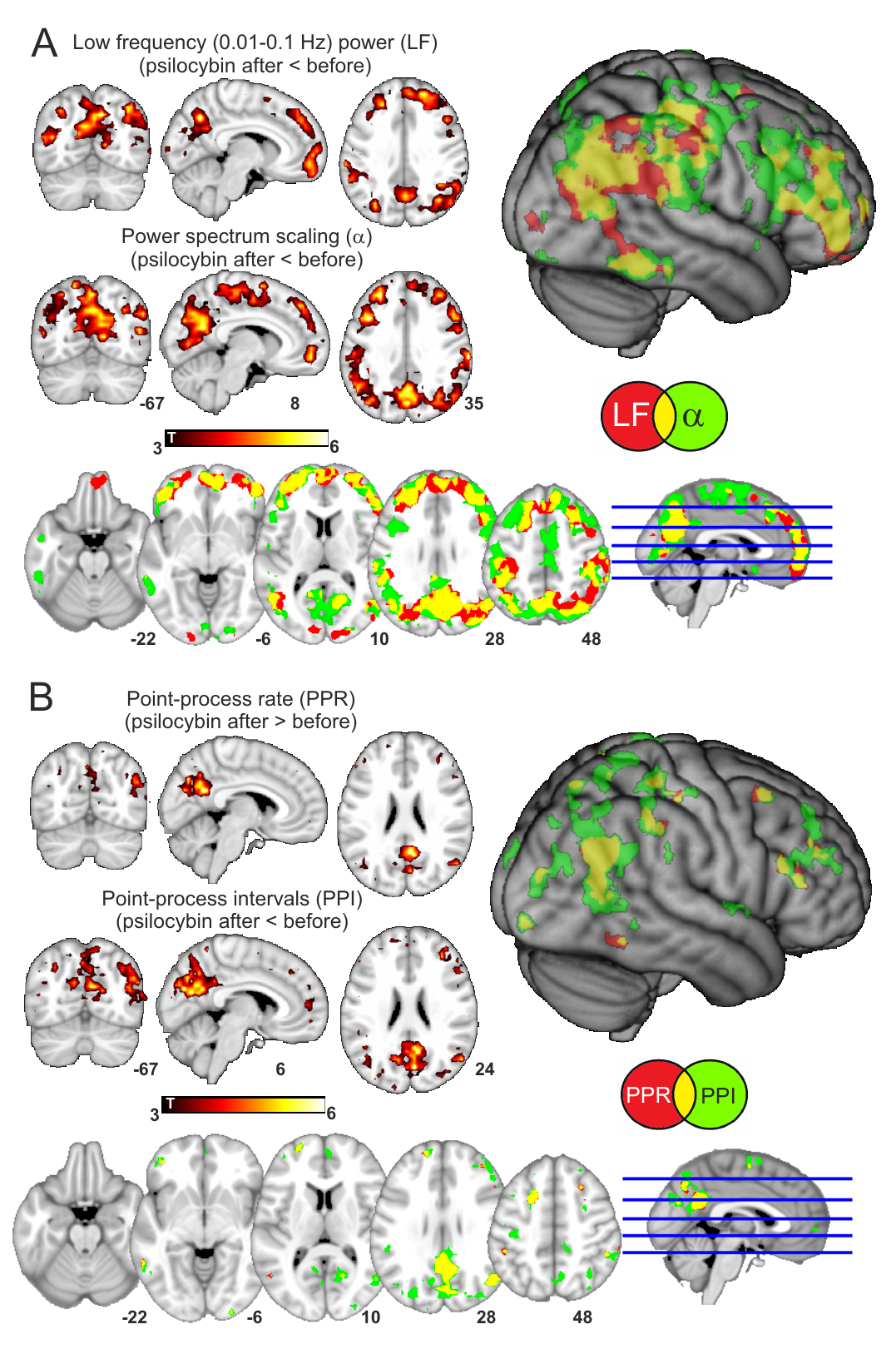,width= 4 truein,clip=true,angle=0} 
\caption{Psilocybin infusion modifies BOLD spectral content in a distributed  fronto-parietal network. A) Maps of statistical significance for decreased low  frequency power (LFP) and power spectrum scaling exponent $\alpha$ after psilocybin 
infusion. Results are shown overlaid separately into an anatomical MNI152 template, 
overlaid together and also rendered together into a 3D anatomical image. Notice 
that in all cases only clusters surviving a threshold of p $<$ 0.05, FWE cluster 
corrected (after passing an uncorrected threshold of p $<$ 0.005) are shown. 
B) Maps of statistical significance of increased power point rate (PPR) and decreased 
point process interval (PPI) after psilocybin infusion (same renderings and statistical 
thresholds as in Panel A).}
\end{figure}

\pagebreak{}

\begin{figure}[ht]
\centering \psfig{figure=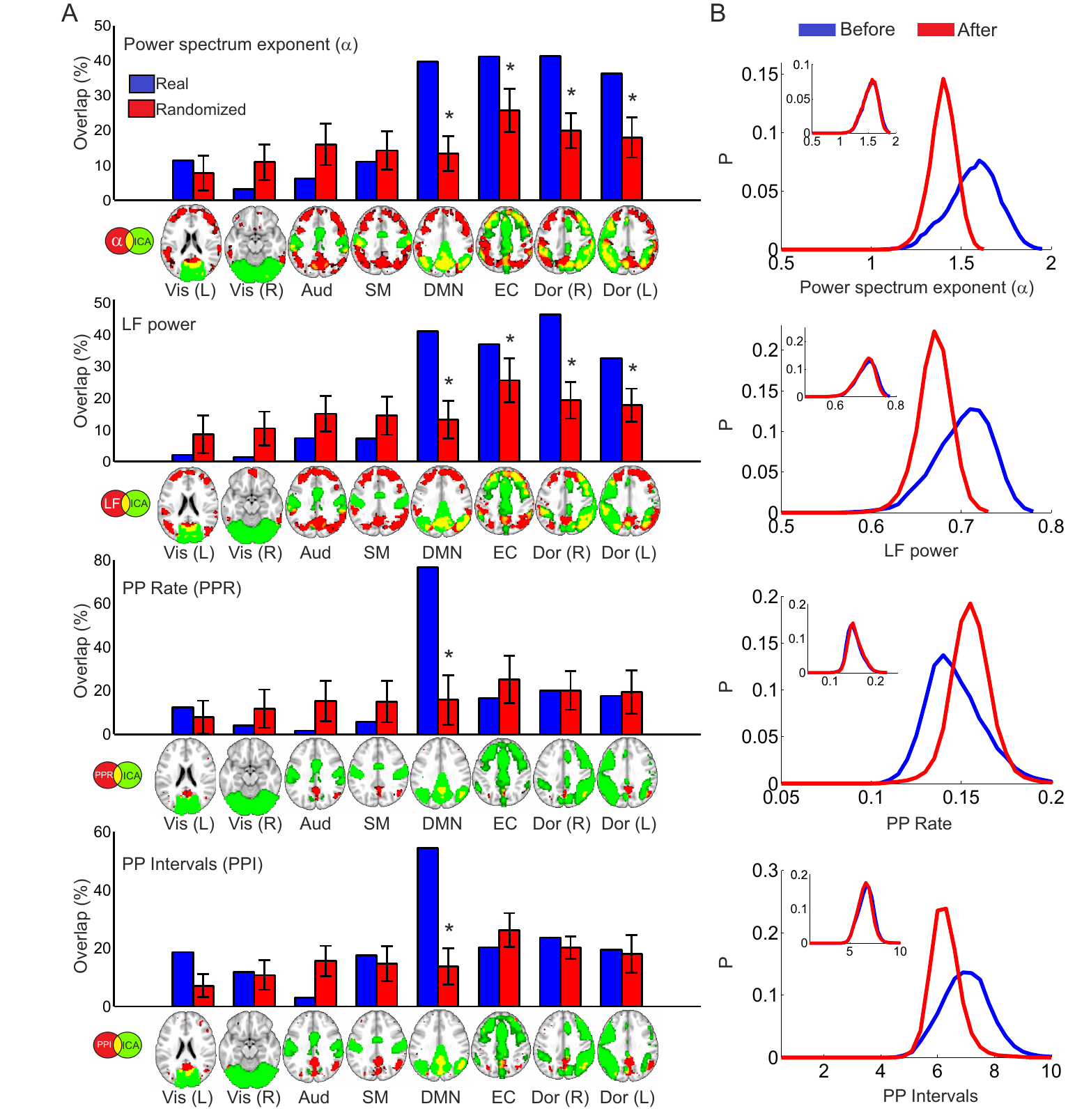,width= 4 truein,clip=true,angle=0} 
\caption{BOLD spectral changes after psilocybin infusion are located 
in higher order brain networks and leave primary sensory areas unaffected. A) Overlap 
between statistical significance maps presented in Fig. 2 and a group of well-established 
cortical RSN (from (Beckmann et al., 2005)) together with the average overlap obtained 
after 100 spatial randomizations preserving first order statistics (image phase 
shuffling). (*) indicates an empirical p-value smaller than 0.05, Bonferroni corrected. 
This p-value is defined as the ratio of instances in which the real maps yield 
a higher overlap than the randomized versions. B) Whole brain grey matter average 
probability distributions for $\alpha$, LFP, PPR and PPI, before and after psilocybin 
infusion. In the inset, the same distributions are shown before and after the placebo 
infusion.}
\end{figure}

\pagebreak{}

\begin{figure}[ht]
\centering \psfig{figure=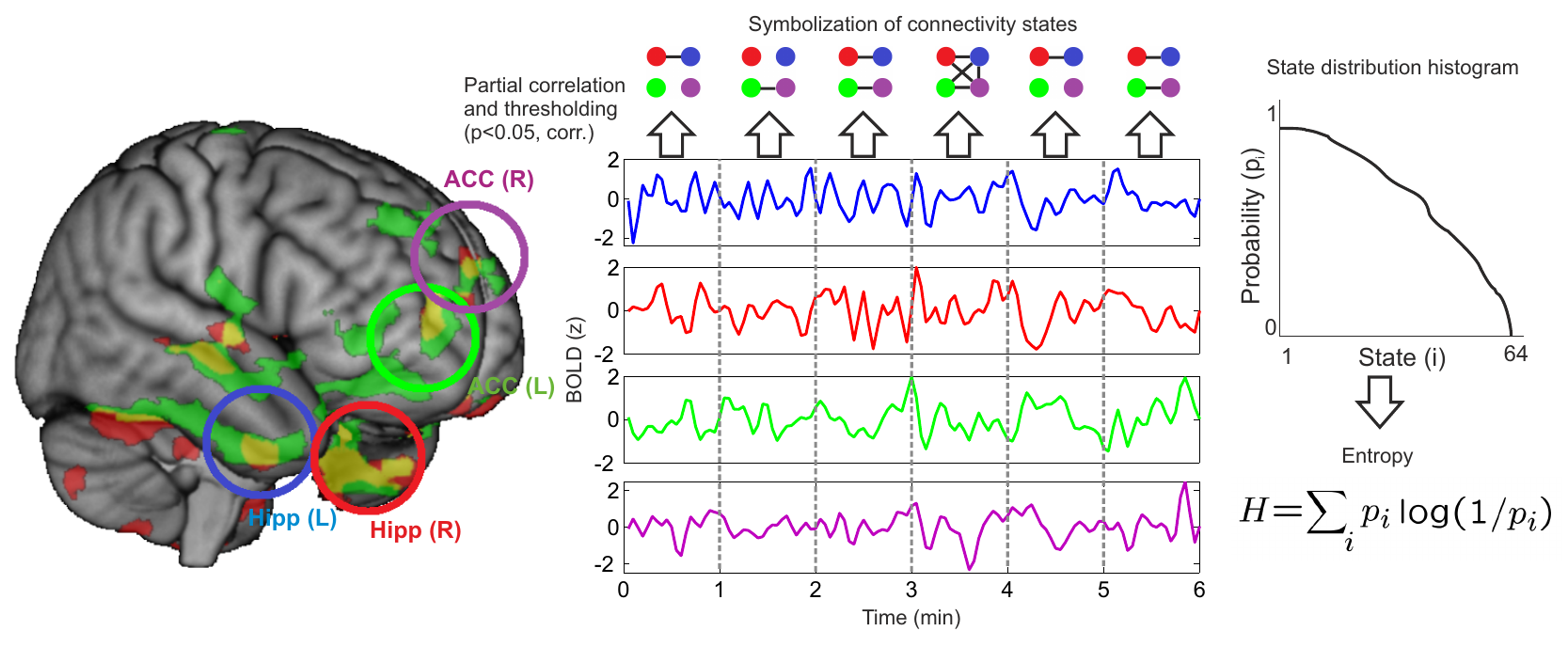,width=5 truein,clip=true,angle=0} 
\caption{Entropy of the dynamical functional connectivity states. Illustration 
of the procedure to estimate the temporal evolution of the collective states (i.e. 
all possible 64 motifs) in the network of regions associated with increased temporal 
variability (bilateral hippocampi and ACC). After selecting the 4 regions of interest 
demonstrating enhanced variability after psilocybin infusion (left and right hippocampus, 
left and right ACC), the partial correlation between all variables is computed 
(including also the mean head displacement time series as a partial regressor). 
After thresholding (with p $<$ 0.05, corrected) a series of up to 64 discrete 
connectivity states are obtained from which the probability distribution can be 
computed. Finally, from this information, histograms of states (provided here as 
an illustration) and their corresponding Shannon's entropy (H) can be computed.}
\end{figure}

\pagebreak{} 

\begin{figure}[ht]
\centering \psfig{figure=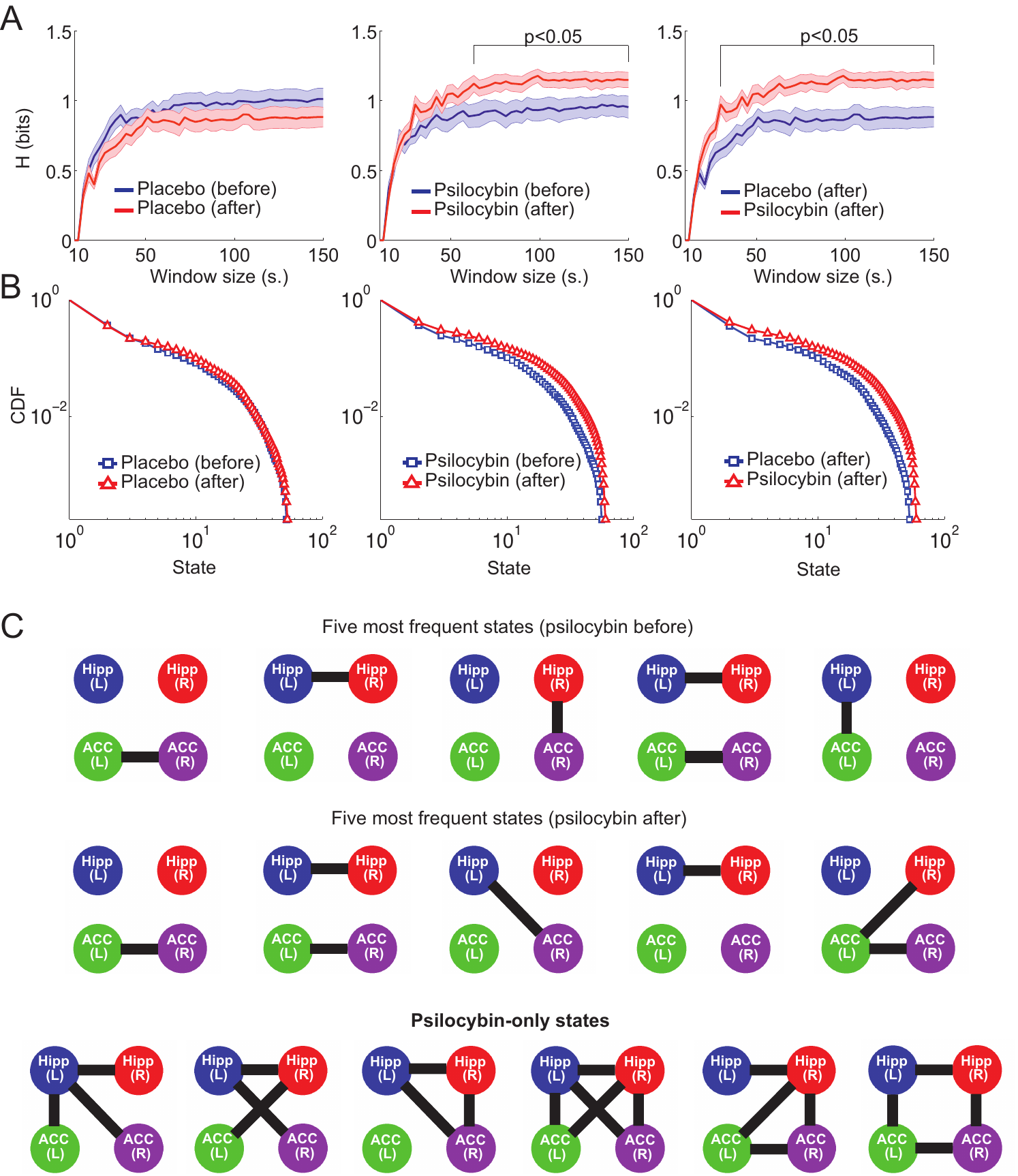,width =4 truein,clip=true,angle=0} 
\caption{Dynamical functional connectivity entropy increased after psilocybin 
infusion. A) Entropy of connectivity states (H, defined following the procedure 
outlined in Fig. 4 and computed at different time window lengths) are plotted for 
the comparisons placebo before vs. after, psilocybin before vs. after and psilocybin 
after vs. placebo after (mean  SEM). B) The probability distributions for the (ranked) 
dynamical functional connectivity states across all conditions. Histograms were 
obtained pooling states across subjects and window sizes. C) First row: five most 
frequent connectivity states before the infusion of psilocybin. Second row: Five 
most frequent states after the infusion of psilocybin. Third row: States observed 
only after the infusion of psilocybin, but absent before the infusion and in the 
placebo condition. In all cases the lines are used to indicate a significant transient 
functional connectivity between two nodes.}
\end{figure}
\pagebreak{}

\end{document}